# Non-extensive analysis of seismicity: application to some seismic sequences of Morocco

Luciano Telesca[1], Taj-Eddine Cherkaoui[2], Mohamed Rouai[3]

[1]National Research Council, Institute of Methodology for Environmental Analysis, Italy

[2]Institut Scientifique, Mohamed V University-Agdal, Rabat, Morocco

[3]GeoTech Research Team, Faculty of Sciences, Meknes University, Meknes, Morocco

*Abstract:* - The magnitude distribution of three seismic sequences occurred in Morocco were investigated by means of the Tsallis-based non-extensive analysis. The non-extensive parameters were estimated by means of the Levenberg-Marquadt nonlinear least square fitting method. It was found that the *q* value could be a good indicator of the complexity of seismic phenomena. Such findings could contribute in better understanding the dynamics of seismicity and suggesting a unifying view of earthquake occurrence.

*Key-Words:* - Non-extensivity; Morocco; magnitude; earthquakes

## 1  Introduction

A log-linear function, called Gutenberg-Richter law (Gutenberg and Richter, 1944), puts the cumulative number of earthquakes (for magnitude $m > M_{th}$) in relationship with the threshold magnitude $M_{th}$. This empirical law is very important because it explains the statistics of the seismicity occurrence, but it was not related with general physical principles. Sotolongo-Costa and Posadas (2004), starting from first principles, developed a general model for earthquake dynamics, in which the Gutenberg-Richter law could be considered as its particular case. In this model not only the irregularities of the fault planes but also the fragments filling the space between them and originated from the breakage of the tectonic plates play an important role in trigger earthquakes. Using the non-extensive Tsallis statistics (Tsallis, 1988), a more realistic magnitude distribution was deduced, providing an excellent fit to the seismicities of several seismic regions (Silva et al., 2006; Telesca, 2010a; Telesca, 2010b; Telesca, 2010c; Telesca and Chen, 2010). In the nonextensive context, Darooneh and Dadashinia (2008) explored the Iranian seismicity, finding that the spatial and temporal distributions between successive earthquakes are described by *q*-exponential function, confirming the Abe's conjecture (Abe et al., 2003) that the sum of the q value in space and time domain is close to 2. Vallianatos (2009), studying the risk function of natural hazards (earthquakes, rockfalls, forest fires, landslides) obtained a relationship between the *b*-value and the non-extensive parameter *q*. The application of non-extensive statistical mechanics to plate tectonics has furnished a *q* value around 1.75, in agreement with the values for the nonextensive parameters found for several seismic areas worldwide (Vallianatos and Sammonds, 2010). Pre-seismic electromagnetic emissions were well-described by the nonextensive model developed for earthquakes, leading to the conclusion that the activation of a single fault is a reduced self-affine image of the regional seismicity and a magnified image of the laboratory seismicity by means of acoustic and electromagnetic emissions (Papadimitriou et al., 2008). Using the Tsallis-like time-dependent entropy a quantitative strategy for monitoring the focal area before earthquake occurrence was proposed (Kalimeri et al., 2008). Interestingly, the non-extensive model for earthquake dynamics also describes the precursory sequence of electromagnetic pulses possibly generated during the activation of a single fault, with similar values of the non-extensive parameter *q*, which can be considered as footprint of self-affine nature of fracture and faulting (Contoyiannis et al., 2010).

## 2  Non-extensive statistics

Complexity in geophysical systems have been becoming to be investigated by means of the non-extensive statistics, developed by Tsallis (1988), which furnishes the theoretical context for analysing some properties of at their nonequilibrium stationary status, like fractality, multifractality, self-similarity, long-range dependencies. In this framework, Sotolongo-Costa and Posadas (2004) and, then, Silva et al. (2006) developed an earthquake fragment-asperity interaction model, in which the earthquake triggering mechanism is given by the interaction between the irregularities of the fault planes with the fragments between them, originated by the local breakage of the tectonic plates, from which the faults are generated. Previously, the irregularity of the profiles of the tectonic plates and fault planes was considered as the main cause of earthquake occurrence; for instance, in De Rubeis et al. (1996) the Gutenberg-Richter law for large earthquakes was obtained assuming that tectonic and fault profiles are Brownian-type and hypothesizing that energy release is proportional to the overlap interval between profiles. The Sotolongo-Costa and Posadas'idea was to combine the irregularities of the fault planes with the distribution of fragments of diverse shape and size between them in order to develop a novel mechanism of triggering earthquakes.

In this model the relative position of fragments filling the space between the irregular fault planes can contribute to the hindering of their relative motion. Then, stress increases until a displacement of one of the asperities, due to the displacement of the hindering fragment, or even its breakage in the point of contact with the fragment leads to a relative displacement of the fault planes of the order of the size $\rho$ of the hindering fragment, producing a liberation of energy $\varepsilon$ (Sotolongo-Costa and Posadas, 2004). Therefore, in this model the mechanism of triggering of earthquakes attributes an important role to the fragments, because they act as roller bearings and hindering entities of the relative motion of the plates until the increasing stress produce their liberation with the subsequent triggering of earthquakes. As large fragments are more difficult to release than small ones, this energy is assumed to be proportional to the volume of the fragment, so that the energy distribution of earthquakes generated by this mechanism can reflect the volumetric distribution of the fragments between plates (Silva et al., 2006). Such assumption revises the former Sotolongo-Costa and Posadas's hypothesis of proportionality between the energy and the linear size of the fragment. Then, the use of nonextensive statistics is a suited tool to describe the volumetric distribution function of the fragments. Applying the maximum entropy principle for the Tsallis entropy (Tsallis, 1988), the Tsallis entropy for our problem is given by

$$T_q = \frac{\int p^q(S)(p(S)^{1-q} - 1)dS}{q-1}, \quad (1)$$

where *p(S)* is the probability of finding a fragment of surface *S* and *q* is a real number (Vilar et al., 2007).

The maximum entropy formulation for Tsallis entropy implies that two conditions have to be introduced: 1) the normalization of *p(S)*, $\int_0^\infty p(S)dS = 1$ ; 2) the ad hoc condition about the *q*-expectation value, $S_q = \langle S \rangle_q = \int_0^\infty S P_q(S)dS$, with escort distribution (Abe, 2003)

$$P_q = \frac{p^q(\sigma)}{\int_0^\infty p^q(\sigma)d\sigma} \quad . (2)$$

Applying the standard method of conditional extremization of the entropy functional $T_q$, we obtain the following expression for the fragment size distribution:

$$p(S) = \left[1 - \frac{(1-q)}{(2-q)}(S - S_q)\right]^{\frac{1}{1-q}} \quad (3)$$

which corresponds to the area distribution for the fragments of the fault planes.

Assuming the energy scale $\varepsilon \sim r^3$, the proportionality between the released energy $\varepsilon$ and $r^3$ becomes $S - S_q = \left(\frac{\varepsilon}{a}\right)^{\frac{2}{3}}$, where *S* scales with $r^2$ and a (the proportionality constant between $\varepsilon$ and $r^3$) has dimensions of volumetric energy density. This scale is in full agreement with the standard theory of seismic moment scaling with rupture length (Lay and Wallace, 1995). Thus, using the transformation $S - S_q = \left(\frac{\varepsilon}{a}\right)^{\frac{2}{3}}$, the following form for the energy distribution function of the earthquakes follows:

$$p(\varepsilon)d\varepsilon = \frac{C_1 \varepsilon^{\frac{1}{3}} d\varepsilon}{\left[1 + C_2 \varepsilon^{\frac{2}{3}}\right]^{1/(q-1)}} \quad . (4)$$

with $C_1 = \dfrac{2}{3a^{\frac{2}{3}}}$ and $C_2 = -\dfrac{(1-q)}{(2-q)a^{\frac{2}{3}}}$. The probability of energy $p(\varepsilon)=n(\varepsilon)/N$, where $n(\varepsilon)$ is the number of earthquakes of energy $\varepsilon$ and $N$ the total number of earthquakes. Therefore, the cumulative number of earthquakes can be calculated as the integral of Eq. (3):

$$\frac{N(\varepsilon > \varepsilon_{th})}{N} = \int_{\varepsilon_{th}}^{\infty} p(\varepsilon) d\varepsilon \quad , (5)$$

Where $N(\varepsilon > \varepsilon_{th})$ is the number of earthquakes with energy larger than the threshold $\varepsilon_{th}$.

Due to the relationship $m \sim log(\varepsilon)$, where $m$ is the magnitude, the following formula for the distribution of the number $N$ of earthquakes whose magnitude $m$ is larger than the threshold $M_{th}$ normalized to the total number of events is obtained:

$$\log\left(\frac{N(m > M_{th})}{N}\right) = \left(\frac{2-q}{1-q}\right) \log\left[1 - \left(\frac{1-q}{2-q}\right)\left(\frac{10^{2M_{th}}}{a^{2/3}}\right)\right] . (6)$$

This not trivial result incorporates the characteristics of non-extensivity into the distribution of earthquakes by magnitude and the Gutenberg-Richter law can be deduced as its particular case.

## 3 Data analysis

In the present paper three seismic Moroccan sequences were investigated: 1) The Al-Hoceima sequence, which includes the series of aftershocks of the strong Al-Hoceima 24 February 2004 earthquake; 2) the West Rif seismicity and 3) the Middle Atlas earthquake series. The time span of all the three seismic series is between 1971 and 2007.

Fig. 1 shows the distribution of the relative number of events (black squares) with magnitude m greater than the threshold $M_{th}$. The fitting was performed by using the software LAB Fit (Silva and Silva, 1999), which uses a nonlinear least square method for performing the curve fitting. There are also shown the curve (red line) fitting the data and representing Eq. 6, along with the 95% confidence band (black dotted lines). There is a nice agreement of Eq. 6 with the data in all the seismic catalogues. In particular, the Al-Hoceima seismicity is characterized by q~1.64 and a~0.9·$10^9$, the West Rif area by q~1.59 and a~2.9·$10^9$, and the Middle Atlas area by q~1.64 and a~0.4·$10^9$. The non-extensive parameters obtained for these sequences are in accordance with those obtained by others in different seismotectonic settings (Vilar et al., 2007). The very good agreement of the Eq. 6 with the data expresses the advantage of nonextensivity model that is based on a physical image that recovers the main characteristics of earthquake dynamics, i.e., the interactions between plates and fragments. The parameter q informs about the scale of these interactions: if q~1, short ranged spatial correlations are present and physical states are close to equilibrium states. As q increases, the physical state goes away from equilibrium states and this implies that the fault planes in the analyzed area are not in equilibrium and more earthquakes can be expected. Regarding the seismic sequences analysed in the present paper, the values obtained for q in each catalogue indicate the presence of long ranged spatial correlations (Sotolongo-Costa and Posadas, 2004).

Since the Al-Hoceima sequence includes the seismic crisis of the Al-Hoceima 24 February 2004 earthquake, we applied the non-extensive statistics on two subsequences extracted from it: one given by the aftershock series of that strong earthquake from February 24, 2004 to May 31, 2004 and the other given by the remaining events of Al-Hoceima seismicity. Fig. 2 shows the normalized number of the events with magnitude m larger than $M_{th}$ for the two subsequences along with the fitting non-extensive curves and their 95% confidence bands. The subsequence of the Al-Hoceima aftershocks is characterized by the non-extensive parameters q~1.65 and a~3.1·$10^9$, while the Al-Hoceima depleted seismic series by q~1.60 and a~0.4·10$^9$. The q-value of the aftershock depleted Al-Hoceima sequence is very close to that of West Rif sequence. Such results suggest that the q-value of the whole Al-Hoceima seismicity is mainly due to the aftershocks; furthermore, they confirm the larger instability of the seismic phenomenon during the aftershock activation given by the larger q-value.

However, the estimate of the q value for each of the analysed areas has to be interpreted as an averaged value, because the seismicity is due to very complex coupling of faults and fault interactions, as well as it is a mixture of locally very intense seismic regions with those characterized by a relatively low seismic activity.

## 4 Conclusion

The nonextensive model for earthquakes was tested on three seismic catalogues in Morocco, and in particular on the aftershock sequence of the strong Al-Hoceima February 24, 2004 event. The model fits well all the catalogues with q values in agreement with those obtained for other seimotectonic settings.

## Acknowledgements

The study presented in this paper was supported by the CNR-CNRST 2010-2011 Bilateral Project "Fractal characterization of seismicity of some active areas of Morocco. Contribution to seismic risk assessment"

.

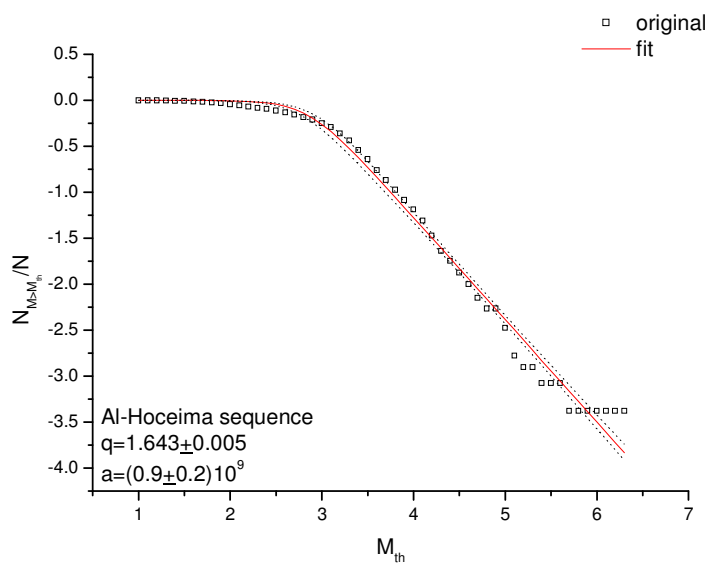

a)

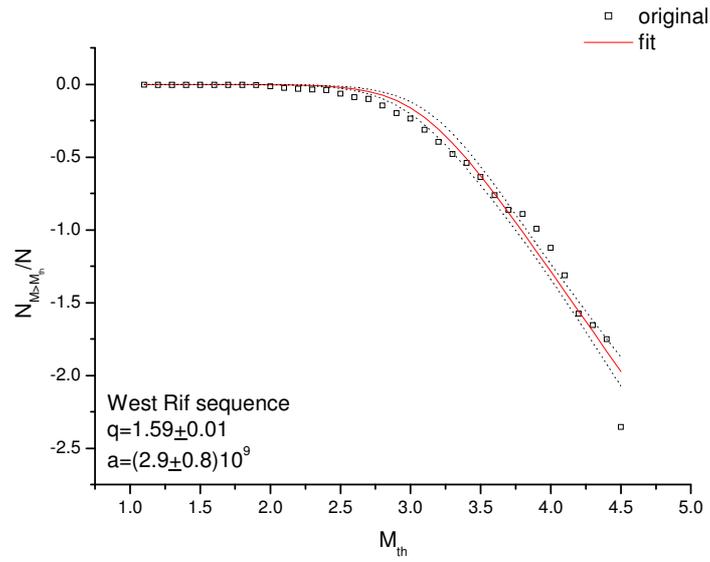

b)

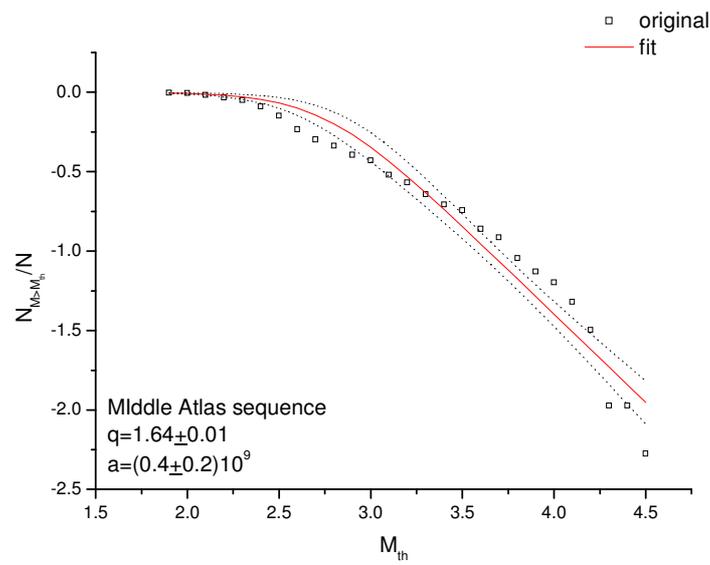

c)

Fig. 1. Non-extensive fit of the normalized cumulative magnitude distribution for the seismic sequences of a) Al-Hoceima, b) West Rif and c) Middle Atlas

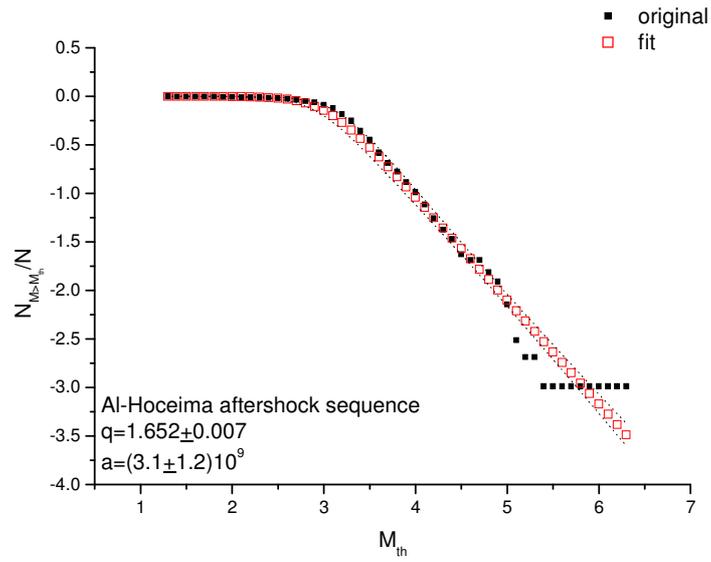

a)

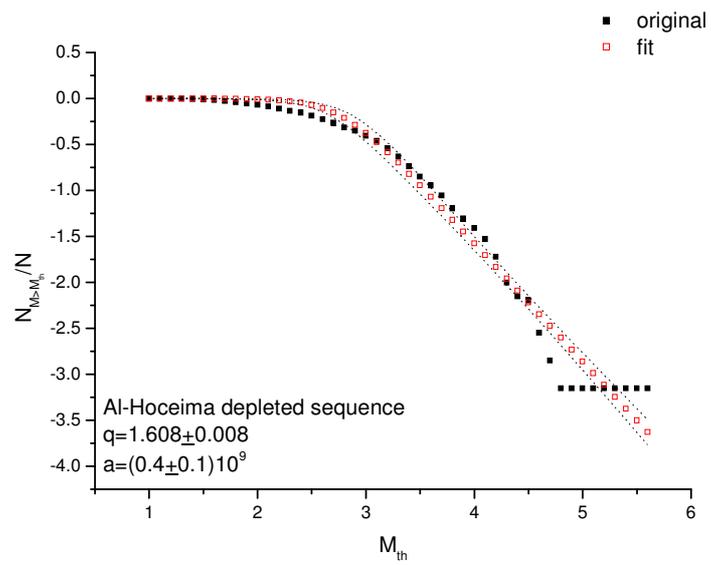

b)

Fig. 2. Non-extensive fit of the normalized cumulative magnitude distribution for the a) Al-Hoceima aftershocks and c) Al-Hoceima aftershock depleted sequence